\documentclass[twocolumn]{aastex631}

\usepackage{comment}
\usepackage{mathtools}
\usepackage{isotope}

\usepackage{booktabs}

\usepackage[hang,flushmargin]{footmisc}
\defcitealias{2019ApJS.243.10P}{MESA V}
\defcitealias{2006ApJ.639.1018W}{W06}
\defcitealias{Yu:18}{YW18}
\defcitealias{Guichandut:2023}{GC23}

\begin{document}

\title{Composition of Radiation-Driven Winds from Type I X-ray Bursts}

\author[0009-0000-2075-1109]{Jason S. Pero}
\affiliation{Department of Physics and Astronomy, University of Kansas, Lawrence, KS 66045, USA; \href{mailto:jpero@ku.edu}{jpero@ku.edu} }

\author[0000-0001-9194-2084]{Nevin N. Weinberg}
\affiliation{Department of Physics, University of Texas at Arlington, Arlington, TX 76019, USA
}
 
\begin{abstract}
Recent NICER observations of photospheric radius expansion (PRE) X-ray bursts reveal absorption features consistent with photospheres enriched in intermediate-mass elements. These features may arise from radiation-driven winds that eject freshly synthesized nuclear ashes, offering a new probe of X-ray bursts and neutron star properties. Motivated by these observations, we use the \texttt{MESA} stellar evolution code to simulate PRE bursts from accretion through the hydrodynamic wind phase. We model a range of ignition depths for both pure helium and mixed hydrogen/helium accretion and explore several prescriptions for convection during burst rise. We find that the wind abundances depend sensitively on both ignition depth and convective treatment, including the efficiency of semiconvective mixing and the prescription used to define convective boundaries. Bursts igniting at column depths $\gtrsim 5\times10^8\textrm{ g cm}^{-2}$ produce ash-enriched winds, with ejecta ranging from intermediate-mass to iron-peak elements depending on ignition depth, accretion composition, and the treatment of convection.
\end{abstract}

\section{\bf I\lowercase{ntroduction}} \label{sec:intro}

Type I X-ray bursts are triggered by the unstable thermonuclear ignition of freshly accreted hydrogen and helium on the surfaces of neutron stars in low-mass X-ray binaries (\citealt{Bildsten:98, Strohmayer:2003, Galloway:2021}). They are observed as bright X-ray flashes lasting seconds to minutes and provide a unique means of probing thermonuclear burning, neutron star (NS) structure, and dense matter physics under extreme conditions. In about 20\% of observed bursts \citep{Galloway:2020}, the radiative luminosity exceeds the local Eddington limit, resulting in a radiation-driven wind that ejects the outer layers of the accreted atmosphere.  During these photospheric radius expansion (PRE) bursts, the photosphere is lifted off the NS surface and pushed outward tens to hundreds of kilometers  (and $\gtrsim10^3$\,km in superexpansion bursts; \citealp{2010AA.520A.81I}). 

In recent years, there has been a renewed effort to use the spectral analysis of X-ray bursts to measure the mass and radii of NSs (see \citealt{Ozel:2016} for a review).  However, because of wind and spectral modeling uncertainties, these measurements may be subject to  systematic errors  \citep{Steiner:10, Boutloukos:10, Suleimanov:11, Miller:13, Miller:16, Medin:16}. In order to  address these uncertainties and  reliably measure NS masses and radii from bursts, we need to better understand the dynamics of the PRE and the composition of  the wind ejecta, and from these compute spectral models that can be compared with observations. 

While the spectra of X-ray bursts are typically well-fit by blackbody models \citep{Galloway:08}, the recent discovery of narrow spectral lines in PRE bursts further motivates the development of improved wind modeling. \citet{Strohmayer:19} find
narrow absorption and emission lines during the PRE phases of four X-ray bursts from 4U 1820-30, including an emission line near 1.0 keV and absorption features near 1.7 and
3.0 keV. They tentatively associate the latter absorption feature with the He-like lines of S XV.  There is a systematic shift in the line energies between pairs of bursts with different PRE strength, which \citet{Strohmayer:19} argue is likely produced by a combination of gravitational redshift and Doppler blueshifts associated with the burst-driven wind. 

\citet{Barra:2025} reanalysed the same four bursts and analysed another eight bursts from 4U 1820-30 observed by {\it NICER}. They detected several significant ($>99.9 \%$ significance) absorption lines, including the 2.97 keV line reported by \citet{Strohmayer:19} and another at 3.4 keV, as well as  emission lines at 1 keV, 2 keV, and 2.4 keV. \citet{Jaisawal:2025} report similar results in their analysis of the same set of bursts (see also \citealt{Yu:2025}).  In addition, they analyze the aftermath emission of a long burst (a superburst) from 4U 1820-30 and detected three absorption lines (see also \citealt{Peng:2025}),  whose energies gradually decrease during the recovery period following the superburst. They argue that the absorption features share a common origin in heavy nuclear ashes enriched with elements like Si, Ar, Ca, or Ti, either from the burst wind or from an accretion flow contaminated by the burst wind.

While \citet{Jaisawal:2025} and \citet{Peng:2025} interpret an absorption line at 3.8 keV following the 4U 1820-30 superburst as coming from an intermediate-mass element in the accretion flow well above the NS surface, \citet{Iaria:2025} interpret it as a gravitationally redshifted Fe absorption line from the NS surface. Using a photoionization absorption model, they measure a redshift of $1 + z =1.72\pm0.05$, which
implies a NS compactness of $R/M = 4.46\pm0.13 \textrm{ km/$M_\sun$}$ ($3.02\pm 0.09$ in dimensionless units). Regardless of which interpretation is correct, the observed spectral features provide a strong incentive to improve our understanding of the dynamics and composition of burst winds.

Early theoretical studies modeled the winds assuming steady-state, optically thick outflows with simplified outer boundary conditions \citep{1983PASJ.35.17E,1983PASJ.35.33K,1985ApJ.289.634Q,1987ApJ.312.700J, Paczynski:86, Nobili:94}. More recently, \citet{Herrera:2020} improved on various aspects of the input physics and numerical techniques to extend the analysis of steady-state solutions and explore the
parameter space in more detail. \citet{Herrera:2023} then  linked these steady-state stellar wind models to time-dependent burst simulations assuming surface atmospheres in hydrostatic equilibrium.  \citet{Guichandut:2021} calculated steady-state models using flux-limited diffusion, which allowed them to account for the transition from optically thick to optically thin regions and thereby extend the steady state solutions out to low optical depths. 

Importantly, the steady-state assumption is often not well satisfied in PRE bursts.  This is because  the PRE lasts only a few seconds in most cases and it takes at least that long for time-independent conditions to be established in the wind \citep{Joss:87}.   Moreover, the wind composition continually varies with time, first ejecting the outer atmospheric layers which contain mostly light, accreted material (H and/or He) and eventually ejecting  the deeper layers that can contain heavy, freshly synthesized ashes of nuclear burning.   The first time-dependent wind models were presented by \citeauthor{Yu:18} (2018; hereafter \citetalias{Yu:18}) who simulated PRE bursts from NSs accreting pure He (as in 4U 1820-30; see \citealt{Cumming:03}). \citeauthor{Guichandut:2023} (2023; hereafter \citetalias{Guichandut:2023}) studied time-dependent winds assuming NSs that accrete a mix of H/He, highlighting the influence of the initial fuel composition and compositional gradients on convection and burst light curves.

In the work presented here, we advance these efforts by systematically exploring how ignition depth, accretion composition, and different prescriptions for convection shape PRE burst wind dynamics and composition.  While \citetalias{Yu:18} also consider a range of ignition depths, they assume pure He accretion whereas we also consider mixed H/He accretion and various convective treatments.  \citetalias{Guichandut:2023} assume mixed H/He accretion, but only consider a single ignition depth and focus primarily on the burst light curve whereas we focus primarily on the wind dynamics and composition. All three studies use the \texttt{MESA} stellar evolution code to model the bursts and winds, with our study taking advantage of several features available in newer versions of \texttt{MESA}.  Most notably, it now includes options to implement a variety of different models of convection (e.g., predictive mixing and convective premixing) whose impact on the burst and wind we explore.  \citetalias{Guichandut:2023} showed that the treatment of convection can significantly influence the predicted burst light curve and we likewise show that it can significantly influence the predicted wind composition. 
 
In Section~\ref{sec:mesa_setup} we describe the \texttt{MESA} setup, including initial models, accretion conditions, and our treatment of the hydrostatic (burst rise) phase and the hydrodynamic (wind) phase.  
Section~\ref{sec:hstat_burst_rise} examines the dynamics of the atmosphere during the hydrostatic phase and Section~\ref{sec:wind_phase} examines the dynamics of the wind during the hydrodynamic phase,  including how these depend on ignition depth and accretion composition.
Section~\ref{sec:wind_composition} describes the wind composition for different ignition depths and convection prescriptions, highlighting the conditions under which heavy-element ashes reach the wind photosphere. We summarize our results and conclude in Section~\ref{sec:summary and conclusions}.

\section{\bf N\lowercase{umerical} M\lowercase{ethods and} M\lowercase{odel} P\lowercase{arameters}}
\label{sec:mesa_setup}
We use the \texttt{MESA} stellar evolution code\footnote{Simulations were compiled using the MESA Software Development Kit, version 24.7.1 available at \cite{richard_townsend_2024_mesasdk}.} (version 23.05.1; Paxton et al. 
\citeyear{Paxton:11},
\citeyear{Paxton:13},
\citeyear{Paxton:15},
\citeyear{Paxton:18},
\citeyear{Paxton:19}, ~\citealt {2023ApJS..265...15J}) to simulate a suite of PRE bursts \footnote{Our \texttt{MESA} inlists, models, and simulation setup are available at \dataset[doi.org:10.5281/zenodo.20713214]{https://doi.org/10.5281/zenodo.20713214}}.  The processes modeled include the accretion of matter, thermonuclear ignition and burning on the NS surface, and the evolution of a radiation-driven wind.

\subsection{Neutron Star and Initial Envelope}

We assume a non-rotating, spherically symmetric NS with a mass $M=1.4M_\odot$ and radius $R=12$ km in Newtonian gravity.  We construct an initial envelope using the \texttt{make\_env} model provided in the \texttt{MESA} test suite.  The initial envelope is  composed of pure ${}^{56}$Fe down to a column depth of $y=10^{12} \textrm{ g cm}^{-2}$, where
\begin{equation}
y(r) = \int_r^\infty \rho \, dr,
\end{equation}
with $r$ and $\rho$ the radial coordinate and mass density, respectively.  Light elements (H/He) accrete on top of this ${}^{56}$Fe layer and eventually the conditions for unstable thermonuclear ignition are reached. The ${}^{56}$Fe layer is meant to serve as a proxy for the inert ashes of burning from previous bursts.  Although the actual composition of burst ashes is not nearly as simple as pure ${}^{56}$Fe (see, e.g., \citealt{Meisel:2019, Zhen:2025}), this should not impact the results of our simulation since our primary aim is to determine the ashes of the freshly synthesized elements ejected in the wind.  While ashes of previous bursts can, in principle, be dredged up and ejected by the wind, our study is not designed to study this possibility.

\subsection{Sources of Pre-Ignition Heating}
\label{subsec:preheating}
There are two potential sources of heating that set the  temperature profile of the accreting atmosphere: deep crustal heating and, if H is present in the accretion material, hot CNO burning. We model the former as a flux $F_{\rm crust} =\dot{m} Q_{\rm crust}$ emanating from below the accreted layers, where $\dot{m}=\dot{M}/4\pi R^2$ is the local mass accretion rate, $\dot{M}$ is the global mass accretion rate, and $Q_{\rm crust}$ is the energy release per nucleon in the crust \citep{Brown:98, Cumming:03, Cumming:06}.  We assume $Q_{\rm crust}=200 \textrm{ keV nucleon}^{-1}$ for all ignition depths and accretion compositions except for the shallowest burst models ($y_b=3\times10^8\textrm{ g cm}^{-2}$) for which we instead assume $Q_{\rm crust}=100 \textrm{ keV nucleon}^{-1}$ due to numerical issues we encountered at higher $Q_{\rm crust}$ (both sets of values are within the theoretical range considered by the studies cited above). In practice, such a flux is implemented in \texttt{MESA} by setting the core luminosity parameter $L_c=4\pi R^2 F_{\rm crust}$.
\begin{deluxetable}{cccccc}
\tablecaption{\texttt{MESA} Ignition Parameters\label{tab:mesa_parameters}}
\tablehead{
\colhead{} & \colhead{} & \multicolumn{2}{c}{\textbf{Pure He}} & \multicolumn{2}{c}{\textbf{Mixed H/He}} \\
\cmidrule(lr){3-4} \cmidrule(lr){5-6}
\colhead{\textup{Label}} & \colhead{$y_\mathrm{b}$}  & \colhead{$\dot{M}$} & \colhead{$L_{\mathrm{c}}$} & \colhead{$\dot{M}$} & \colhead{$L_{\text{c}}$} \\ 
\colhead{} & \colhead{($\textrm{g cm}^{-2}$)} & \colhead{($\dot{M}_{\rm Edd}$)} & \colhead{($L_\odot$)} & \colhead{($\dot{M}_{\rm Edd}$)} & \colhead{($L_\odot$)}}
\startdata
y3e8 & $3.0\times 10^8$ & 0.35 & 20.2  & 0.05 &  1.8\\
y5e8 & $5.0\times 10^8$ & 0.11 & 12.9 & 0.03 & 1.9 \\
y1e9 & $1.0 \times 10^9$  & 0.07 & 7.8 & 0.02 & 1.3 \\
y5e9 & $5.0\times 10^9$ & 0.03 & 3.3 & 0.01 & 0.7 
\enddata
\end{deluxetable}

The flux $F_{\rm H}$ from stable H burning via the hot CNO cycle is self-consistently computed by \texttt{MESA}, which  tracks the depletion of H (and formation of He) with depth through the accreted layers.  As described in \citet{Cumming:00} and \citetalias{Guichandut:2023}, the H is completely depleted at a column depth 
\begin{equation}\label{eq:yd}
    y_\mathrm{d}\simeq 5.4\times 10^8\,\mathrm{g\,cm^{-2}}\!{\left(\frac{\dot{M}}{0.1\dot{M}_\mathrm{Edd}}\right)}\!\!{\left(\frac{0.01}{Z_\mathrm{CNO}}\right)}\!\!{\left(\frac{X_\mathrm{0}}{0.7}\right),}
\end{equation}
where $X_0$ and $Z_\mathrm{CNO}$ are the initial H and CNO nuclei mass fractions, respectively, and we scale $\dot{M}$ by the Eddington accretion rate  $\dot{M}_{\rm Edd}=8\pi R m_p c/ (1+X_0) \sigma_T$ corresponding to a neutron star with $R= 12\textrm{ km}$ and $X_\mathrm{0}=0.7$. In all our simulations, the burst ignition depth $y_{\rm b} > y_{\rm d}$ and thus all our bursts ignite in a pure He layer (aside from trace amounts of CNO).

\subsection{Burst Ignition Depths}
\label{subsec:ignition depths}

We consider four ignition depths in our study  $y_\mathrm{b}=\{3, 5, 10, 50\}\times 10^8\,\mathrm{g\,cm^{-2}}$, which we will designate as models $\{$y3e8, y5e8, y1e9, y5e9$\}$, respectively. For each ignition depth, we consider a pure He accretion model and a mixed H/He accretion model.  For the latter, we assume initial $^1$H, $^4$He, and $^{12}$C mass fractions of $X=0.7$, $Y=0.28$, $Z=0.02$, respectively.  As in \citetalias{Guichandut:2023}, we assume for simplicity that $^{12}$C is the only metal accreted. Since the CNO abundances adjust to the equilibrium ratio of $^{14}$O and $^{15}$O in the hot CNO cycle, this simplification has little effect on the burst ignition and nucleosynthesis. We demonstrate this quantitatively in Section~\ref{subsec:impact_burning_and_solar} by simulating a burst with a more realistic solar abundance accretion. 

To obtain models that ignite at these four specific depths, we vary the accretion rate $\dot{M}$ (which in turn also sets $L_c$, as described above).  The values of $\dot{M}$ and $L_c$ for each model are given in Table~\ref{tab:mesa_parameters}.  For a given ignition depth, the mixed H/He models have a lower $L_c$ and $\dot{M}$ because their accreting atmospheres are kept warm not just by deep crustal heating but also by the hot CNO cycle.  The two shallower depths (y3e8, y5e8), are representative of short duration bursts (which are by far the most common type of bursts observed) whose total durations are a few seconds to tens of seconds, while the two larger ignition depths (y1e9, y5e9) are representative of intermediate duration bursts which last a few minutes to tens of minutes. Short and long duration bursts are found to occur in systems accreting both pure He and mixed H/He \citep{Galloway:2021}.   
\\
\subsection{Reaction Network}

We use \texttt{MESA}'s \texttt{approx21.net} reaction network throughout our calculations (including during the wind phase). The \texttt{approx21.net} network builds on the $\alpha$-chain backbone by incorporating simplified models for steady-state H burning (including the PP chain and CNO cycle), carbon and oxygen burning reactions ($\isotope[12]{C}+\isotope[12]{C}$, $\isotope[12]{C}+\isotope[16]{O}$, and $\isotope[16]{O}+\isotope[16]{O}$), and certain aspects of photodisintegration involving \(^{54}\mathrm{Fe}\), as outlined in \cite{1978ApJ.225.1021W}. It also introduces the isotopes \(^{56}\mathrm{Cr}\) and \(^{56}\mathrm{Fe}\). Nuclear reaction rates are from JINA REACLIB \citep{Cyburt2010}, NACRE \citep{Angulo1999}, and additional tabulated weak reaction rates \citep{Fuller1985ApJ, Oda1994ADNDT, Langanke2000}. Screening is included via the prescription of \cite{Chugunov2007}. Thermal neutrino loss rates are from \cite{Itoh1996}. Incorporating larger nuclear reaction networks, capable of capturing extended $rp$-process burning, presented numerical challenges and are left for future work.

\subsection{Baseline Treatment of Convection}\label{subsec:reference_model_config}
Our baseline treatment of convection assumes the Ledoux criterion, which accounts for stabilizing gradients in both temperature and composition. We use the Henyey formulation \citep{1965ApJ...142..841H}, which includes a self-consistent treatment of radiative losses from convective elements and is better suited for inefficient convection. Thermohaline mixing is included using the \cite{1980AA.91.175K} prescription with a thermohaline coefficient $\alpha_{\mathrm{th}}$=2, enabling mixing in regions where the mean molecular weight decreases inward. The semiconvection efficiency parameter is set to the \texttt{MESA} default value of $\alpha_{\mathrm{sc}}$=0.

This simplified, baseline treatment of convection may miss important physics, and the \texttt{MESA} documentation recommends enabling additional convection prescriptions, particularly in the presence of composition gradients. In Section~\ref{sec:wind_composition}, we consider different treatments of convection and assess the extent to which these can ultimately influence the wind structure, evolution, and composition. In general, we find that while the overall wind structure and timescales are relatively insensitive to the treatment of convection,, the wind's composition can vary greatly depending on the treatment of convection.

\subsection{Transition to Hydrodynamic Wind Phase}
The unstable ignition of helium marks the formal onset of the burst and the resulting energy release quickly heats up the overlying layers.  Within $\simeq 1\textrm{ s}$, the radiative luminosity in the top-most layers of the atmosphere  exceeds the local Eddington limit
\begin{equation}
    L_{\rm Edd}(X, T) = \dfrac{4\pi c GM}{\kappa(X, T)}. 
    \label{eq:Ledd}
\end{equation}
The opacity $\kappa(X, T)$ is dominated by electron scattering and depends on composition (primarily the H mass fraction $X$) and temperature due to relativistic (Klein-Nishina) corrections.  As the radiation diffuses upward into shallower, cooler layers, $\kappa$ increases and the radiative luminosity transitions from being sub-Eddington ($L_{\rm rad} <L_{\rm Edd}$) to super-Eddington ($L_{\rm rad} >L_{\rm Edd}$). This marks the onset of the wind stage and the moment we switch from using \texttt{MESA}'s hydrostatic solver to its hydrodynamic solver. In practice, we implement this by modifying the \texttt{extras\_check\_model} routine in \texttt{run\_star\_extras.f90} to stop the hydrostatic simulation when $L_\mathrm{rad}/L_\mathrm{Edd} \geq 0.9995$ in the outermost layers. This threshold is chosen so that the model never formally reaches or exceeds the Eddington limit under the hydrostatic solver, which anyway begins to experience numerical difficulties when this condition is met. 

The subsequent wind phase is modeled as a nearly purely radiative outflow by disabling convective mixing (turning off \texttt{MESA}'s implementation of mixing length theory), turning off accretion, and activating the hydrodynamic solver to track the wind dynamics and composition. During the hydrostatic phase, the top boundary is set at an optical depth of $\tau=100$ to prevent numerical instabilities that can arise in shallower layers. During the wind phase, however, we set the top boundary at $\tau=1 $ to capture the regions near the photosphere.

\begin{figure*}[!ht]
    \centering
    \includegraphics[scale=1]{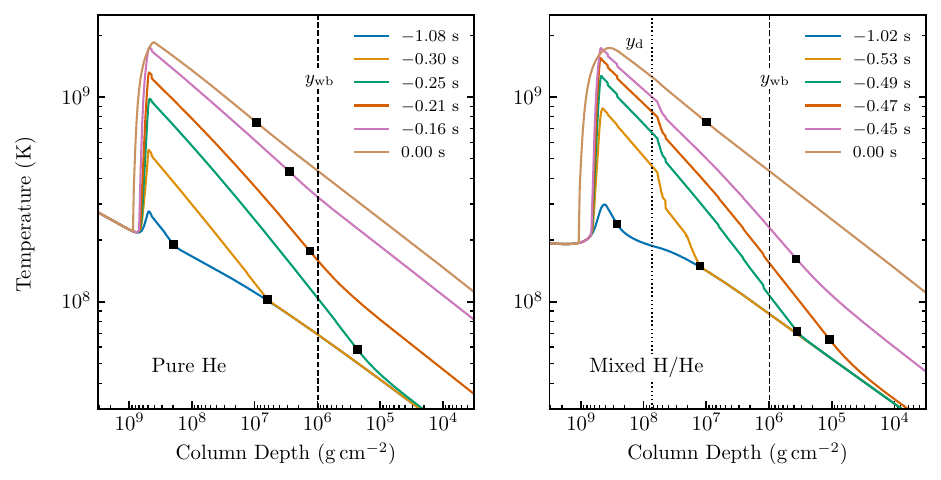}
    \caption{Temperature as a function of column depth for ignition at $y_b=5\times10^8 \textrm{ g cm}^{-2}$ assuming pure He accretion (left panel) and mixed H/He accretion (right panel). The different line colors correspond to different times during the burst rise, with time labeled in seconds and $t=0$ corresponding to the start of the radiation-driven wind. Squares indicate the top of the convection zone $y_c$ and the dashed vertical lines mark the maximum depth of the wind base $y_\mathrm{wb}$ during the subsequent wind phase (the wind will eject material above this line). The dotted line in the mixed H/He panel  is the hydrogen depletion depth $y_\mathrm{d}$.}
    \label{fig:fig1}
\end{figure*}

Sharp gradients in the H mass fractions in the pre-wind atmosphere created numerical challenges when simulating the winds and led to density inversions that disrupted the \texttt{MESA} integration. To address this, we applied the same smoothing spline technique used by \citetalias{Guichandut:2023} to soften the composition jumps before running the mixed H/He wind simulations.

\section{Hydrostatic Burst Rise} \label{sec:hstat_burst_rise}
The evolution of the hydrostatic envelope during the rise phase of a burst plays a critical role in determining the extent of convective mixing and the subsequent composition of material ejected in the wind. In this section, we examine the thermal and convective response of the envelope from the onset of unstable nuclear burning to the point just before wind launch.  In Section~\ref{subsec:effects_ignition_depth_on_convection}, we analyze how the depth of ignition influences the extent of convective mixing, comparing pure He and mixed H/He accretion models. In Section~\ref{subsec:split_convection}, we describe the prevalence of split convection zones in the simulations.  Finally, in Section~\ref{subsec:realistic_mixed}, we evaluate the sensitivity of convective mixing to the detailed elemental composition of the accreted material by comparing models with simplified and realistic solar abundances. The results presented in this section and the next use our baseline treatment of convection (see Section~\ref{subsec:reference_model_config}); we consider different treatments of convection in Section~\ref{subsec:dependence_convection}.

\begin{figure*}
    \centering
    \includegraphics[scale=1]{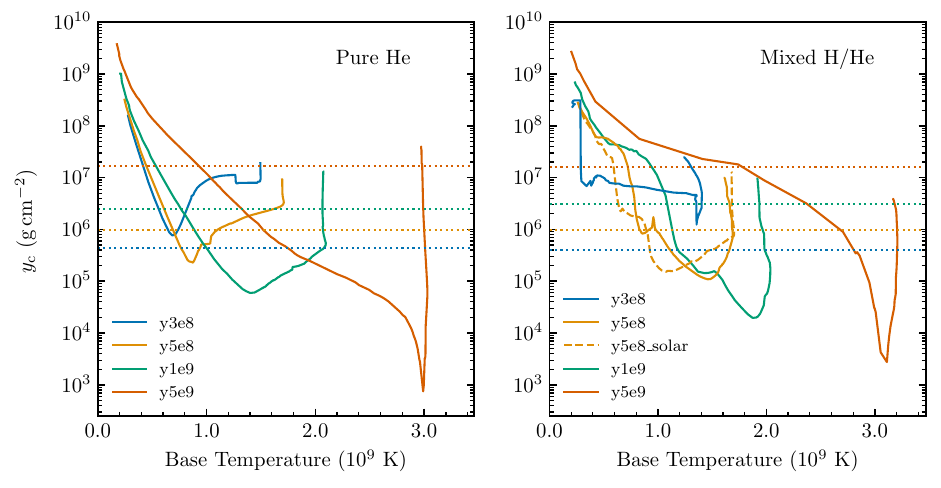}
    \caption{Evolution of the top of the convection zone $y_c$ as a function of base temperature $T_b$ during the burst rise (before wind launch) for models with pure He accretion (left panel) and mixed H/He accretion (right panel). The horizontal dotted lines denote the wind base $y_{\rm wb}$, color coded by the respective models.}
    \label{fig:nw_fig6}
\end{figure*}

\subsection{Effects of Ignition Depth and Accretion Composition on the Burst Rise} \label{subsec:effects_ignition_depth_on_convection}

Figure~\ref{fig:fig1} shows the evolution of the temperature as a function of column depth during the burst rise for the y5e8 models assuming pure He accretion (left panel) and mixed H/He accretion (right panel). The time labels indicate seconds before $t=0$, defined as the start of the wind (the moment when $L_{\rm rad}$ first exceeds $L_{\rm Edd}$ in the outer atmosphere). Squares mark the top of the convection zone, while the dashed vertical line denotes the maximum inward extent of the wind base $y_{\mathrm{wb}}$ during the PRE phase; material above this depth is ejected in the wind.  We define the column depth of the wind base  as
\begin{equation}
    y_{\rm wb}(t) = \frac{M_{\rm ej}(t)}{4\pi R^2}
\end{equation} 
where the total mass ejected at time $t>0$ is given by $M_{\rm ej}(t) =\int_0^{t} \dot{M}_{\rm w}(t^\prime)dt^\prime$. Here the mass-loss rate $\dot{M}_{\rm w} = 4\pi r^2\rho v$, where $\rho$ and $v$ are the density and velocity of the wind at radius $r$ and time $t$.

In both models, the convection zone initially extends outward along an adiabat before eventually retreating as radiative diffusion carries an increasing fraction of the flux, consistent with the hydrostatic burst-rise evolution described by \citeauthor{2006ApJ.639.1018W} (2006; hereafter \citetalias{2006ApJ.639.1018W}) and \citetalias{Yu:18}. The column depth at the top of the convection zone, $y_c$, reaches a minimum value that is smaller than $y_{\mathrm{wb}}$, implying that the wind will entrain some of the convectively mixed ashes of thermonuclear burning. In each panel of Figure~\ref{fig:fig1}, the rightmost black square marks this minimum $y_c$.

As noted in Section~\ref{subsec:preheating},  the accretion rate is slow enough that all the H in the mixed H/He models burns via the hot CNO cycle prior to He ignition. Therefore, their ignition occurs in nearly pure He layers just like the pure He accretion models.  Nonetheless, there are two notable differences in their convective evolution.  First, as can be seen in Figure~\ref{fig:fig1}, the mixed H/He models have steep temperature gradients just shallower than the hydrogen depletion depth $y_\mathrm{d}$ (see Equation~(\ref{eq:yd})) due to sharp gradients in the mean molecular weight between the heavy ashes and the light H-rich material.  This compositional stratification tends to inhibit the outward growth of the convective zone until the material becomes sufficiently hot to overcome the  buoyancy barrier, as discussed in \citetalias{2006ApJ.639.1018W}.
Second, for models y5e8 and y1e9, the minimum $y_c$ is somewhat smaller in the mixed H/He models compared to the pure He models (see also Figure~\ref{fig:nw_fig6} discussed next). As described in \citetalias{Guichandut:2023}, this is due to the collision between the convection zone and the overlying H layer which brings in fresh protons below $y_{\rm d}$,
causing a rapid injection of energy and a further expansion of the convection zone. 

Figure \ref{fig:nw_fig6} shows how the top of the convection zone $y_\mathrm{c}$ evolves as a function of base temperature $T_\mathrm{b}$ during the burst rise up until the wind is launched. The left panel shows models with pure He accretion and the right panel shows models with mixed H/He accretion. Each curve corresponds to a different ignition column depth $y_\mathrm{b}$, and the maximum value of $y_{\rm wb}$ is indicated  by the horizontal dotted lines.  In both pure He and mixed H/He accretion, the convection zone initially extends outward from the base of the burning layer as the temperature rises. The extent of this outward growth depends strongly on the ignition depth and the composition of the accreted material. In the shallowest ignition models ($y_b=3\times 10^8\, \mathrm{g\,cm^{-2}}$), the convective boundary $y_c$ never reaches $y_{\rm wb}$. However, for deeper ignition depths ($y_b\ge 5\times 10^8 \textrm{ g cm}^{-2}$), the convection zone extends well past $y_{\rm wb}$, indicating that ashes of nuclear burning can be ejected during the subsequent wind phase.

\begin{figure*}[!ht]
\centering
\includegraphics[scale=1]{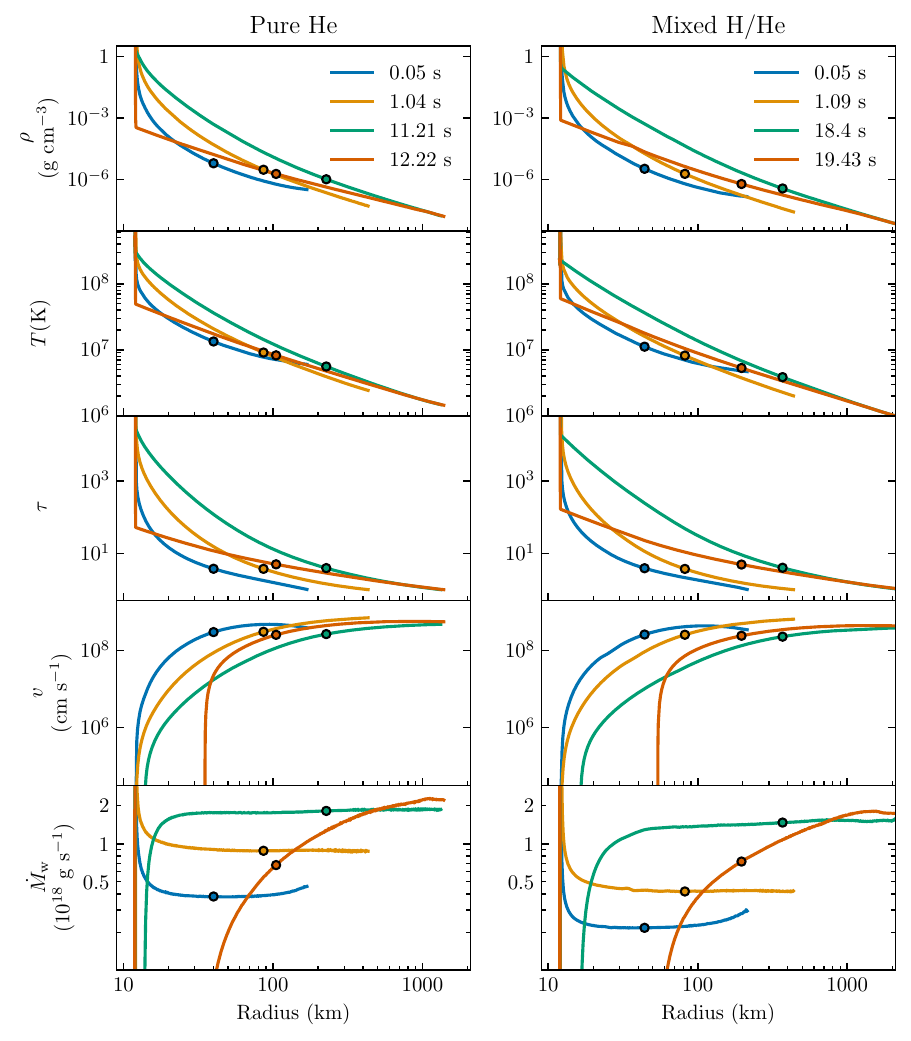}
\caption{Radial profiles of the wind structure at different times for ignition at $y_b=5\times 10^8\textrm{ g cm}^{-2}$ assuming pure He accretion (left panels) and mixed H/He accretion (right panels). The panels show, from top to bottom, the mass density $\rho$, temperature $T$, optical depth $\tau$, wind velocity $v$, and mass loss rate $\dot{M}_{\rm w}$. Each curve is terminated at the location where the optical depth $\tau = 1$. The circles indicate the location of the photosphere $r_{\rm ph}$. }
    \label{fig:radial_profiles_of_wind}
\end{figure*}

The distinct shape of the mixed H/He y3e8 curve in the right panel, which shows a relatively shallow and gradual slope in $y_\mathrm{c}$ as $T_{\rm b}$ increases, can be explained by the shallow ignition depth and the timing of heat transport during the burst rise. For shallow bursts like y3e8, the location of peak temperature does not initially occur at the base of the accreted layer. Instead, it starts slightly above the base and slowly moves downward as burning progresses. This delays the development of a steep inward temperature gradient near $y_\mathrm{b}$ and prevents rapid expansion of the convection zone. By contrast, for deeper ignitions, such as those shown in Figure~\ref{fig:fig1}, the temperature rises sharply at $y_\mathrm{b}$ and the top of the convection zone moves outward to lower column depths more rapidly with increasing $T_{\rm b}$.

\subsection{Split Convection} \label{subsec:split_convection}
Split convection zones are regions within a convective envelope characterized by the temporary interruption of convective flow by stable, radiative layers. These zones arise from a variety of physical conditions, including local changes in opacity, composition gradients, or complex heat transport mechanisms. 

\citetalias{Guichandut:2023} investigated split convection zones, finding that they emerge as a natural consequence of the rapid interplay between nuclear burning and convection, especially following the collision of the convection zone with the H-rich radiative layers at $y<y_{\rm d}$. This process, driven by rapid energy injection from proton captures, causes the convection to split into multiple zones interspersed by tiny radiative gaps. However, the authors highlighted that the resultant complexity is highly sensitive to the chosen convection prescription and spatial resolution, noting that this splitting might be an artifact of simplified one-dimensional assumptions and that simulations only converge when boundary effects like overshoot mixing are explicitly modeled.

We find evidence of split convection across all of our simulations, with a prevalence that depends on which treatment of convection is adopted (it is more prevalent in the predictive mixing and CPM treatments, which we describe in Section~\ref{subsec:dependence_convection}). Models computed at higher spatial resolution also exhibit a greater number of split convective zones and wider radiative gaps, consistent with the convergence test results reported by \citetalias{Guichandut:2023}.

\subsection{More Realistic Mixed Accretion Composition} \label{subsec:realistic_mixed}

As discussed in Section~\ref{subsec:ignition depths}, the mixed H/He models assume initial $^1$H, $^4$He, and $^{12}$C mass fractions of $X=0.7$, $Y=0.28$, $Z=0.02$, respectively.  To examine the impact of this simplified treatment of metal abundances, we ran a y5e8 model with a more realistic solar accretion abundance based on the values of \citeauthor{2009M&PSA..72.5154L} (2009; we call this model y5e8\textunderscore solar).  We find that for the same $\dot{M}$, both ignite at $y_b=5\times10^8\textrm{ g cm}^{-2}$, but that the top of the convection zone of  the y5e8\textunderscore solar model reaches slightly shallower column depths, placing it slightly farther above $y_{\rm wb}$ (compare the y5e8 and y5e8\textunderscore solar curves in Figure~\ref{fig:nw_fig6}). However, as we will show in Section~\ref{sec:wind_composition} (see Figure~\ref{fig:mass_frac_vs_mass_num_solar}), this has only a very modest effect on the composition of the ashes ejected in the wind.

\section{Wind Dynamics} \label{sec:wind_phase}

When the radiative luminosity $L_{\rm rad}$ in the atmosphere first exceeds the local Eddington limit $L_\mathrm{Edd}$, a radiation-driven wind is launched.\footnote{The condition $L_{\rm rad}(y) > L_{\rm Edd}(y)$ occurs first in the outermost layers and then over time moves inward to larger $y$ (see Figure~5 in \citetalias{Yu:18}).}  As described in Section \ref{sec:mesa_setup}, at that moment we stop the hydrostatic simulations and use the last hydrostatic profile as the starting model for the  hydrodynamic wind simulations. 

\begin{figure*}[!ht]
    \centering
    \includegraphics[scale=1]{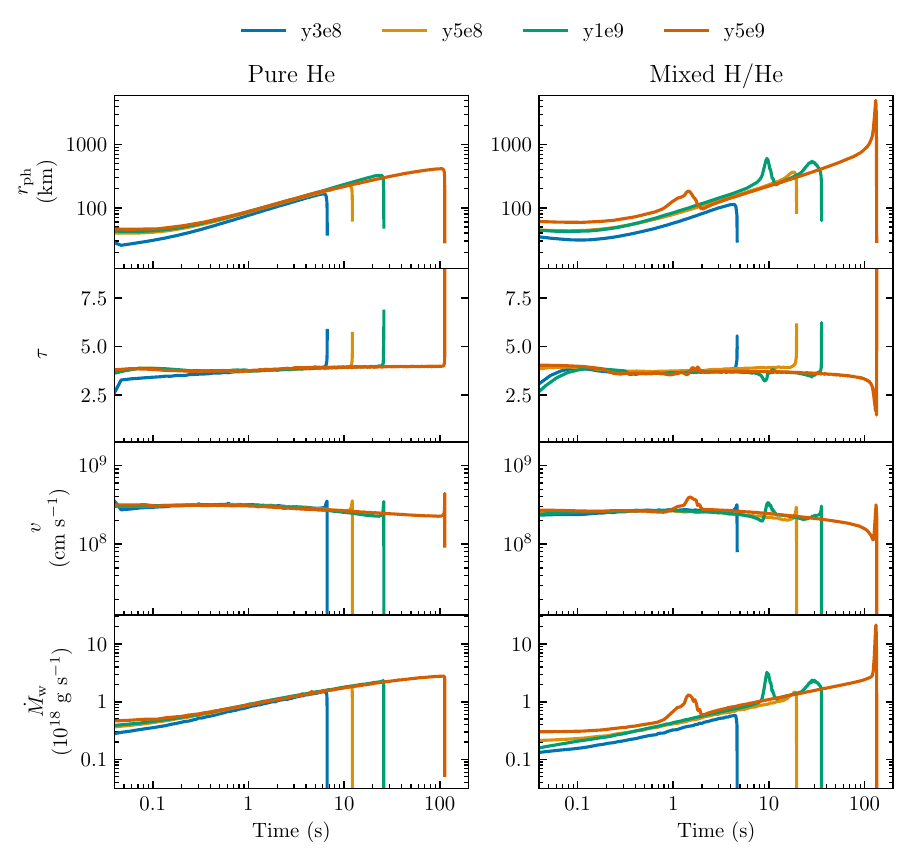}
    \caption{Wind properties evaluated at the photosphere $r_{\rm ph}$ as a function of time for the four burst ignition depths (listed at the top) assuming pure He accretion (left panels) and mixed H/He accretion (right panels).  The panels show, from top to bottom, the evolution of the photospheric radius $r_{\rm ph}$, the optical depth $\tau$, the wind velocity $v$, and the mass loss rate $\dot{M}_{\rm w}$.}
    \label{fig:wind_properties_at_photosphere}
\end{figure*}

In this section, we focus on the evolution of the wind structure and the time-dependence of its global parameters, examining how they depend on ignition depth and accretion composition (we will describe the wind composition in  Section~\ref{sec:wind_composition}). In Section~\ref{subsec:time evolution of y5e8 wind}, we describe results from the two representative y5e8 models (pure He and mixed H/He).  Although the overall wind structure and evolution are similar across the various models,  there are some important differences that we describe in Section~\ref{subsec:wind trends}.

\subsection{Time Evolution of the Wind Profiles}
\label{subsec:time evolution of y5e8 wind}

In Figure \ref{fig:radial_profiles_of_wind} we show the evolution of the wind structure of the y5e8 models assuming pure He accretion (left panels) and mixed H/He accretion (right panels). The different panels show from top to bottom, respectively, the radial profiles of density $\rho$, temperature $T$, optical depth $\tau$, wind velocity $v$, and mass loss rate $\dot{M}_{\rm w} =4\pi r^2\rho v$.  The curves are each truncated at optical depth $\tau=1$, and the location of the photosphere $r_{\rm ph}$ is marked by an open circle. Following \citetalias{Yu:18}, we define  $r_{\rm ph}$ as the radius where $\sigma T^4 = L_{\rm rad}/4\pi r^2$; as can be seen in the optical depth profiles, $r_{\rm ph}$ typically corresponds to $\tau\approx 3-5$ (similar to estimates in \citealt{Quinn:85, Paczynski:86}). In both the pure He and mixed H/He accretion models, the wind gradually strengthens over time, with increases in $\rho$, $T$, $\tau$, and $\dot{M}_{\rm w}$, consistent with the results of \citetalias{Yu:18}. 

For the pure He accretion y5e8 model, we find that the wind settles into a steady state at time $t \simeq 1 \textrm{ s}$, with only modest changes over the next $\approx 10 \textrm{ s}$. During this phase, the mass-loss rate reaches a peak value of $\dot{M}_{\rm w} \simeq 1.8\times10^{18}\textrm{ g s}^{-1}$ and the photospheric radius reaches a maximum value of \( \mathrm{r_{ph} \simeq 230 \, \mathrm{km}} \).
By \( t \simeq  12 \, \text{s} \), the atmosphere has cooled and $L_{\rm rad}$ decreases below $L_{\rm Edd}$  causing a rapid decline in $\rho$, $T$, $v$, and $\dot{M}_{\rm w}$ that begins near the NS surface.

The wind profiles for the mixed H/He accretion y5e8 model are overall similar to that of the pure He accretion model. One notable difference is that mixed H/He model evolves somewhat more slowly, as evident from the longer timescales, with the profiles reaching steady state at $t \simeq 1 \textrm{ s}$ and remaining relatively unchanged for the next $\approx 18 \textrm{ s}$. The maximum mass-loss rate is slightly lower too, with $\dot{M}_{\rm w} \simeq 1.5 \times 10^{18} \textrm{ g s}^{-1}$, and the maximum photospheric radius is somewhat larger, reaching $r_{\rm ph} \simeq 370 \textrm{ km}$ (see also the top panel of Figure \ref{fig:wind_properties_at_photosphere} discussed below). 

The reason the wind from the mixed H/He model lasts longer and is more spatially extended than that of the pure He model is likely because of the composition dependence of the electron scattering opacity $\kappa(X, T)$ and thus the Eddington limit $L_{\rm Edd}(X, T)$ (see Equation~(\ref{eq:Ledd})). When H is present, the larger number of free electrons per nucleus increases $\kappa$ which lowers $L_{\rm Edd}$. As illustrated in Figure 9 of \citetalias{2006ApJ.639.1018W}, this means that for the same $L_\mathrm{rad}$, a mixed H/He atmosphere will be super-Eddington over a broader range of depths than a pure He atmosphere. In an evolving burst, this lower $L_\mathrm{Edd}$ allows the mixed case to remain super-Eddington for longer as the atmosphere cools after the burst rise, sustaining an extended photosphere and wind after the pure He case has already dropped below its Eddington limit. An additional factor that may contribute to the mixed H/He models' longer lasting, more extended winds is the enhanced energy release from H burning after the collision between the convection zone and the overlying H layer.

\subsection{Dependence of Wind Profiles on Ignition Depth and Accretion Composition}
\label{subsec:wind trends}

We now describe how key properties of the wind depend on ignition depth as well as accretion composition.   Figure \ref{fig:wind_properties_at_photosphere} shows the time evolution of $r_{\rm ph}$, $\tau(r_{\rm ph})$, $v(r_{\rm ph})$, and $\dot{M}_{\rm w}(r_{\rm ph})$ for bursts at four different ignition column depths ($y_\mathrm{b}=3\times 10^8$, $5\times 10^8$, $1\times 10^9$, and $5\times 10^9$ g $\mathrm{cm^{-2}}$), comparing models with pure He accretion (left panels) and mixed H/He accretion (right panels).  The time evolution of these quantities offers insight into the structure and dynamics of the wind and its sensitivity to both ignition depth and accretion composition.

Overall, we find that the deeper the ignition, the longer the wind duration and the larger the peak $r_{\rm ph}$. The duration is a few seconds for $y_b=3\times10^8 \textrm{ g cm}^{-2}$ and over 100 seconds for $y_b=5\times10^9 \textrm{ g cm}^{-2}$. This is because deeper ignitions release more energy and thus have longer lasting and more powerful winds. 

While these wind parameters (especially duration) are sensitive to ignition depth, we see that they have a relatively mild dependence on accretion composition. Consistent with the y5e8 models discussed above, the peak value of $r_{\rm ph}$ ($\dot{M}_{\rm w}$) is somewhat larger (smaller) for the mixed H/He accretion models, although nearly all have maximum radii of $r_\mathrm{ph} \approx 200-400 \textrm{ km}$ (maximum mass loss rates of $\dot{M}_{\rm w} \approx 1-3 \times10^{18}\textrm{ g s}^{-1}$).  The one exception is the y5e9 mixed H/He accretion model which reaches an especially large  radius $r_\mathrm{ph} \gtrsim 1000 \textrm{ km}$. 

The optical depth at the photosphere is a nearly constant value of $\tau \simeq$ 3.5 independent of time, ignition depth, and accretion composition.  This is consistent with with the arguments presented in \citet{Quinn:85} and \citet{Paczynski:86}. The apparent spikes above $\tau\sim$3.5 are not physical; they occur when the wind subsides and the location of $r_\mathrm{ph}$ becomes difficult to resolve numerically.

\begin{figure*}[!ht]
    \centering
    \includegraphics[scale=1]{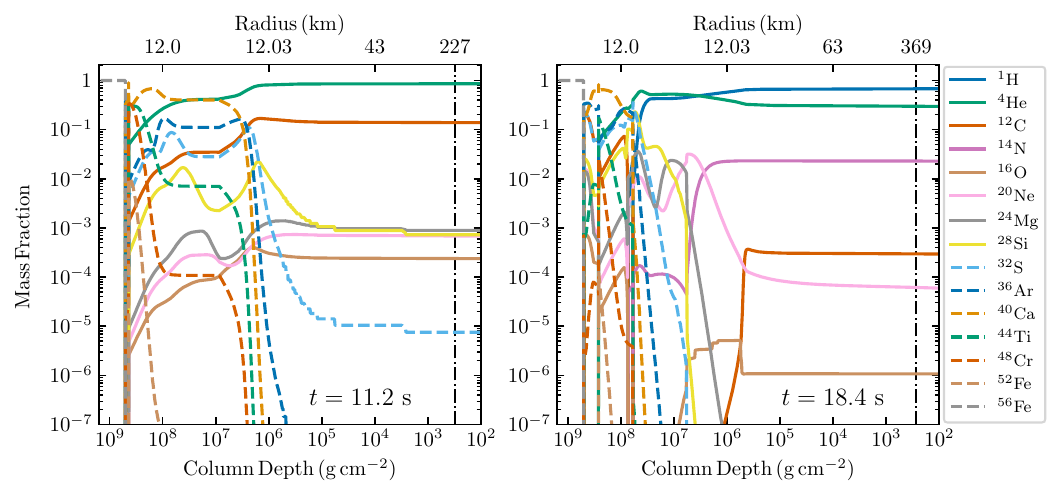}
    \caption{Mass fraction of elements during the maximum radial extent of the wind as a function of column depth (bottom axis) and radius (top axis) for our baseline y5e8 model. The left panel is the pure He accretion model at $t=11.2 \textrm{ s}$, while the right panel is the mixed H/He accretion model at $t=18.4 \textrm{ s}$ (corresponding to each model's time of maximum $r_{\rm ph}$). The line styles are shown in the right panel key and represent the same elements in both panels. The vertical dashed-dotted line marks the location of the photosphere $r_{\rm ph}$.}
    \label{fig:fig9_y5e8}
\end{figure*}

\section{Wind Composition} \label{sec:wind_composition}

In this section we explore how the composition of material ejected during the wind depends on the accretion composition, ignition depth, and the scheme used to model convection during the hydrostatic burst rise.  In Section~\ref{subsec:influence_of_initial_fuel} we show the wind compositions for our baseline y5e8 models with pure He accretion and mixed H/He accretion. In Section~\ref{subsec:wind_dependence_semiconvection} we examine how the y5e8 results depend on the choice of the semiconvective mixing efficiency parameter $\alpha_{\rm sc}$ and in Section~\ref{subsec:impact_ignition_depth} we show how the wind composition depends on ignition depth $y_b$ (for various values of $\alpha_{\rm sc}$). In Section~\ref{subsec:dependence_convection} we explore different convection prescriptions and their influence on the wind composition.  Finally, in Section~\ref{subsec:impact_burning_and_solar}, we test the role of active nuclear burning during the wind phase and assess the impact of the accreted metal abundance by comparing the carbon-only case (our baseline model) with accretion of solar metal abundances. Together, these analyses provide a multifaceted  view of how physical effects and numerical choices can shape the nucleosynthetic signatures observable in burst winds.

\begin{figure*}[!ht]
    \centering
    \includegraphics[scale=1]{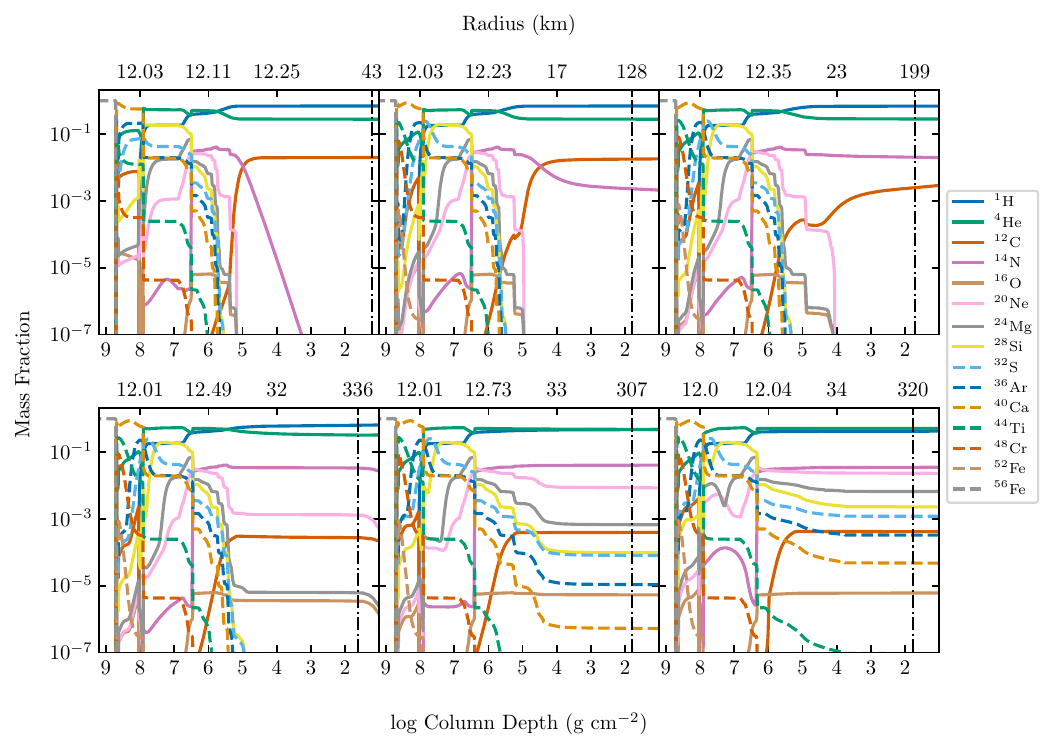}
    \caption{Evolution of the wind composition profile for the y5e8 model with mixed H/He accretion assuming a semiconvection efficiency   parameter $\alpha_{sc}$ = 0.5. The top row (left to right) shows the profiles at $t=0.05$, $t=2$, and $t=4 \textrm{ s}$ and the bottom row (left to right) shows them at $t=6$, $t=10$, and $t=16 \textrm{ s}$. The vertical dashed-dotted line indicates the location of the photosphere.}
    \label{fig:fig9_compare6_y5e8}
\end{figure*}

\subsection{Dependence on Accretion Composition}\label{subsec:influence_of_initial_fuel}

Figure \ref{fig:fig9_y5e8} shows the wind composition for our baseline y5e8 models, plotted as a function of column depth (bottom axis) and radius (top axis). The composition profiles are taken at $t \approx 11.2$ s for the pure He accretion model (left panel) and $t \approx 18.4$ s for the mixed H/He accretion model (right panel).  These times correspond to the maximum radial extent of the wind (maximum $r_{\rm ph}$) in each case. The vertical dash-dotted line indicates the location of the photosphere $r_{\rm ph}$. 

In the pure He accretion model, the wind's photosphere is rich in intermediate mass elements with prominent mass fractions of \isotope[12]{C}, \isotope[16]{O}, \isotope[20]{Ne}, \isotope[24]{Mg}, \isotope[28]{Si}, and \isotope[32]{S} which were synthesized during the burst. The mass fraction of \isotope[4]{He} decreases rapidly for $y\gtrsim 10^8\textrm{ g cm}^{-2}$, indicating efficient helium consumption in the burning layer. As expected, \isotope[1]{H} is absent due to the lack of hydrogen in the accreted material. The composition from the photosphere at $r_{\rm ph}\simeq 230 \textrm{ km}$ down to $r\simeq 20 \textrm{ km}$ is relatively uniform and dominated by intermediate mass elements. This reflects the pre-wind composition of the outer  atmosphere, whose layers are stratified by the convection zone's retreat and which the wind gradually exposes.

In the mixed H/He accretion model, the wind is primarily composed of the accreted H/He that resides in the top layers of the atmosphere (at $y< y_{\rm d}$) as well as trace amounts of \isotope[12]{C} and \isotope[16]{O} associated with pre-burst hot CNO burning. Given that the pure He and mixed H/He models have fairly similar ash-rich compositions near the burning layer (at $y\gtrsim 10^8\textrm{ g cm}^{-2}$), their different wind composition shows that convection during the burst rise is less efficient at bringing heavy elements above the wind base $y_\mathrm{wb}$ in the mixed H/He model.  This occurs because the lighter elements in the mixed H/He model enhance the atmosphere's compositional stratification and thus suppress convection and convective mixing. This highlights the influence of accretion composition on the wind composition, with potential implications for the spectral signatures of heavy elements in PRE bursts. However, as we will show in Sections ~\ref{subsec:wind_dependence_semiconvection}, \ref{subsec:impact_ignition_depth}, and \ref{subsec:dependence_convection}, mixed H/He models can eject heavier elements in the wind depending on the semiconvective mixing efficiency, ignition depth, and  how convection is modeled, respectively.

\subsection{Impact of Semiconvective Mixing}\label{subsec:wind_dependence_semiconvection}

In our baseline y5e8 models shown in Figure~\ref{fig:fig9_y5e8}, we neglected the possibility of semiconvective mixing during the hydrostatic burst rise, i.e., we set $\alpha_{\rm sc}=0$ in \texttt{MESA}. In the next section, we consider a range of values for $\alpha_{\rm sc}$ and present results for the four ignition depths $y_b$. In this section, we begin with a detailed comparison for the y5e8 mixed H/He accretion model.

In Figure \ref{fig:fig9_compare6_y5e8} we show the time evolution of the wind composition for $\alpha_{\mathrm{sc}} = 0.5$ (we also show results for $\alpha_{\mathrm{sc}}=0.1$ in the next section). Each panel displays the elemental mass fractions as a function of column depth (bottom axis) and radius (top axis) at a different stage of the wind. The top row (left to right) shows  it at $t = 0.05$, 2, and 4 s, and the bottom row shows it at $t = 6$, 10, and 16 s. The vertical dash-dotted line in each panel marks the location of the photosphere.

Near the onset of the wind ($t = 0.05$ s), the outflow is just beginning to develop. The composition above the photosphere is still dominated by unburned light elements, primarily \isotope[1]{H} and \isotope[4]{He}. A significant amount of \isotope[12]{C}, originating from the accreted material, is also ejected at this early time. The heavier nuclei produced by helium burning, such as \isotope[16]{O} and \isotope[20]{Ne}, remain confined to deeper layers at this early stage.

By $t = 2$ and $4\textrm{ s}$, the wind has intensified and the photosphere has moved out to $r_{\rm ph}\simeq 100-200 \textrm{ km}$. Some intermediate mass elements that were synthesized during the burst, such as 
\isotope[20]{Ne} and \isotope[24]{Mg}, begin to be lifted off the NS surface.  At $t = 6\textrm{ s}$, ashes appear at the photosphere in small but increasing amounts, indicating that the wind has  peeled away the outer atmosphere and is beginning to expose the deeper layers.  

By $t=10\textrm{ s}$, and even more so at $t=25\textrm{ s}$,  the wind contains appreciable mass fractions of \isotope[28]{Si}, \isotope[32]{S}, \isotope[36]{Ar}, and \isotope[40]{Ca} at the photosphere (although \isotope[1]{H} and \isotope[4]{He} always dominate the composition). These intermediate mass elements are products of $\alpha$-capture, and their presence signals that semiconvection (with $\alpha_{\rm sc}=0.5$) has enabled significant mixing between the deeper burning layers near $y_b$ and the wind base at $y_{\rm wb}$. 

\begin{figure}[!t]
    \centering
    \includegraphics[scale=1]{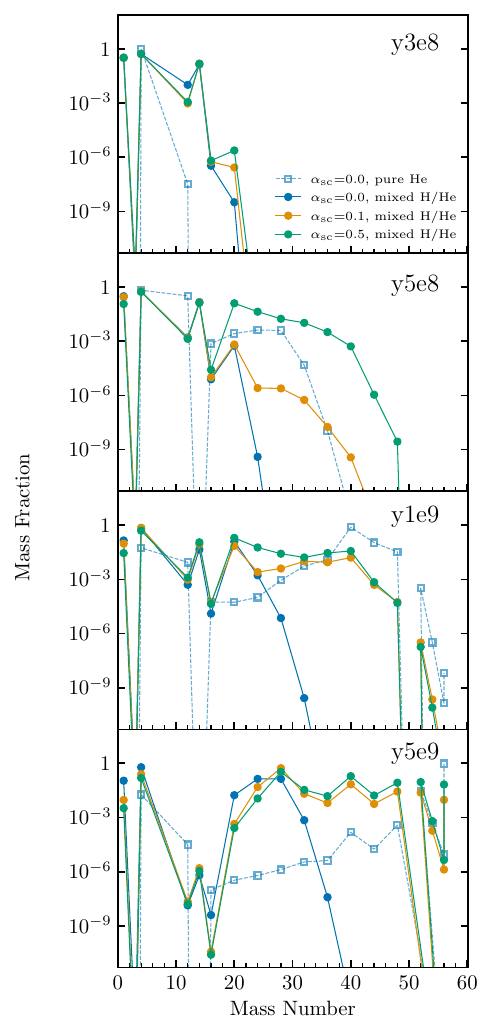}
    \caption{Composition of the wind at the photosphere ($r=r_{\rm ph}$) at the time of maximum $r_{\rm ph}$ for the pure He models (squares) and the mixed H/He models (circles).  Each panel corresponds to a different ignition depth: $y_b = 3 \times 10^8$, $5 \times 10^8$, $1 \times 10^9$, and $5 \times 10^9 \textrm{ g cm}^{-2}$ (top to bottom). For the mixed H/He models, we shows results for three values of the semiconvection efficiency parameter $\alpha_{\rm sc}=0, 0.1, 0.5$. The pure He accretion models are all with  $\alpha_{\rm sc}=0$.}
    \label{fig:mass_frac_vs_mass_num_asc}
\end{figure}

\subsection{Impact of Ignition Depth}\label{subsec:impact_ignition_depth}

We now describe how the wind composition depends on ignition depth $y_b$.  Figure \ref{fig:mass_frac_vs_mass_num_asc} shows the composition at the photosphere at the time of maximum $r_{\rm ph}$ for the mixed H/He accretion models with $y_b = 3 \times 10^8$, $5 \times 10^8$, $1 \times 10^9$, and $5 \times 10^9 \textrm{ g cm}^{-2}$ (top to bottom). We show results for three values of the semiconvection  efficiency parameter $\alpha_{\mathrm{sc}} = 0.0$, 0.1, and 0.5. This range of values is motivated by \citet{Kaiser:2020}, who in their study of the relative importance of convective uncertainties in massive stars considered $\alpha_{\rm sc}=0.004$ and $0.4$ as representing slow and fast semiconvection.

For the shallowest ignition depth (y3e8 models), the wind composition remains dominated by H and He and closely resembles the initial accreted abundances, with barely any ashes ejected regardless of $\alpha_{\rm sc}$. This result reflects the limited  extent of convection $y_c$ compared to the depth of the wind base $y_{\rm wb}$ (see the right panel of Figure \ref{fig:nw_fig6}).

For the y5e8 models, the effects of semiconvective mixing become apparent as $\alpha_{\mathrm{sc}}$ increases. For $\alpha_{\mathrm{sc}} = 0.1$, and especially for $\alpha_{\mathrm{sc}} = 0.5$ (see also Section~\ref{subsec:wind_dependence_semiconvection}), the mass fractions of intermediate mass elements (mass numbers $30-40$)  are clearly enhanced, indicating mixing of ashes from the burning layer. While light elements still dominate, the appearance of these heavier nuclei in the photosphere suggests the possibility of compositionally enriched outflows in short duration (i.e., shallow ignition depth), mixed H/He PRE bursts.

At the deepest ignition depths (y1e9 and y5e9 models), the dependence on $\alpha_{\rm sc}$ is especially striking.   In the absence of semiconvection ($\alpha_{\mathrm{sc}} = 0.0$), the wind of the y1e9 models remains H and He rich, with only traces of heavier elements. However, for $\alpha_{\mathrm{sc}} = 0.1$ and 0.5, the wind composition is highly enriched in ashes of advanced $\alpha$-capture burning, including \isotope[36]{Ar}, \isotope[40]{Ca}, and \isotope[44]{Ti}. For the y5e9 models,  iron group nuclei such as \isotope[52]{Fe}, \isotope[56]{Fe}, and \isotope[56]{Ni} are present in the photosphere even for $\alpha_{\rm sc}=0.1$.

These results show that for the mixed H/He accretion models, ignition depth and semiconvection mixing efficiency significantly impact the chemical makeup of the wind. Deeper ignition depths naturally produce more processed material and shallower $y_\mathrm{c}$ (see Figure~\ref{fig:nw_fig6}), but the extent to which that material reaches the wind's photosphere depends strongly on the efficiency of semiconvective mixing during the hydrostatic pre-wind phase. 

Figure~\ref{fig:mass_frac_vs_mass_num_asc} also shows results for the pure He models at the same four values of $y_b$ (assuming $\alpha_{\rm sc}=0$). Similar to the mixed H/He models, the composition is dominated by the accretion composition for the y3e8 model and becomes increasingly rich in heavier elements with increasing $y_b$. The key difference is that for the pure He models, ashes are ejected even with $\alpha_{\rm sc}=0$.  This is because there is no overlying H layer to suppress convective mixing (see discussion at end of Section~\ref{subsec:influence_of_initial_fuel}).

\begin{figure}[!t]
    \centering
    \includegraphics[scale=1]{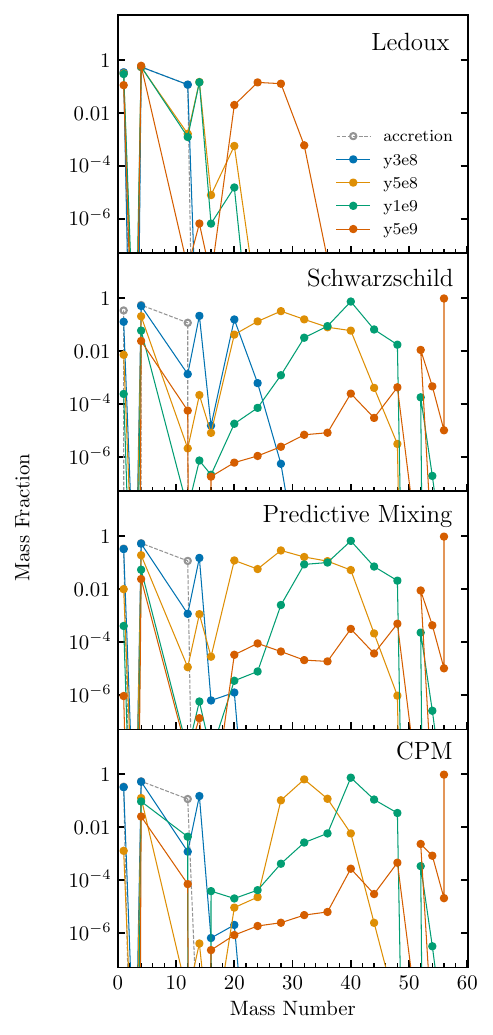}
    \caption{Composition of the wind at the photosphere ($r=r_{\rm ph}$) at the time of maximum $r_{\rm ph}$ assuming mixed H/He accretion.  Each panel corresponds to a different treatment of convection: the Ledoux criterion, the Schwarschild criterion, predictive mixing, and CPM (top to bottom). Within each panel, we show results for the four ignition depths: $y_b = 3 \times 10^8$, $5 \times 10^8$, $1 \times 10^9$, and $5 \times 10^9 \textrm{ g cm}^{-2}$. The gray open circles indicate the accreted composition.}
    \label{fig:mass_frac_vs_mass_num_all}
\end{figure}

\subsection{Dependence on How Convection is Modeled}\label{subsec:dependence_convection}

In our baseline \texttt{MESA} setup, we use the Ledoux criterion to model convection during the burst rise. According to the Ledoux criterion, convective boundaries are located where the discriminant $\nabla_{\rm rad}-\nabla_{\rm L}$ changes sign; here $\nabla_{\rm rad}$ is the radiative temperature gradient and $\nabla_{\rm L}$ is the Ledoux temperature gradient, which takes into account the effect of composition gradients (see Equation~(11) of \citealt{Paxton:13}). In this section, we explore how the following three alternative \texttt{MESA} treatments of convection impact the wind composition: the Schwarzschild criterion, predictive mixing, and convective premixing (CPM). We briefly describe each of these and refer the reader to \citet{Paxton:13, Paxton:18, Paxton:19} for detailed descriptions.

According to the Schwarzschild criterion, convective boundaries are located where the discriminant $\nabla_{\rm rad}-\nabla_{\rm ad}$ changes sign, where $\nabla_{\rm ad}$ is the adiabatic temperature gradient; thus, unlike the Ledoux criterion, it ignores the effects of composition gradients.  As described in \citet{Paxton:18}, an issue with using the Ledoux or Schwarschild criteria in \texttt{MESA} is that the location of a convective boundary is not uniquely determined but instead depends on the mixing history near the boundary. As a result, it is possible that $\nabla_{\rm rad} > \nabla_{\rm ad}$ on the convective side of the boundary, which is physically inconsistent with the local mixing length theory requirement that $\nabla_{\rm rad} = \nabla_{\rm ad}$ on the convective side of the boundary. To resolve this problem,  \citet{Paxton:18} introduced the predictive mixing scheme which improves on \texttt{MESA}'s previous algorithms by allowing a convective region to expand during a time step until its boundary satisfies $\nabla_{\rm rad}=\nabla_{\rm ad}$ on the convective side. However, \citeauthor{Paxton:18} (2018; see also \citealt{Paxton:19}) found certain cases that continue to exhibit $\nabla_{\rm rad} > \nabla_{\rm ad}$ on the convective side of the boundary even with predictive mixing.  The CPM scheme introduced in \citet{Paxton:19} is designed to address these issues and better track convective boundaries by directly modifying abundances each time step through an iterative series of mixings over a shifting window of computational cells.  

Predictive mixing in \texttt{MESA} is governed by a set of parameters that control how the scheme operates within predefined convection zones (\texttt{core}, \texttt{shell}, and \texttt{surf}). In our models, predictive mixing functioned as intended only when both the \texttt{shell} (convective shell) and \texttt{surf} (surface convection zone) regions were explicitly enabled. Once these zones are activated, it is also necessary to specify the minimum superadiabaticity required for predictive mixing to operate in each region. The \texttt{MESA} documentation recommends adopting a minimum superadiabaticity of at least 0.005 for core He-burning models in order to prevent artificial splitting of the core convection zone and the occurrence of core breathing pulses. Motivated by this guidance, we adopted a threshold of 0.005 for the \texttt{shell} zone and a slightly lower value of 0.001 for the \texttt{surf} zone in our burst models with predictive mixing.

In Figure \ref{fig:mass_frac_vs_mass_num_all}, we show the wind composition at the photosphere at the time of maximum $r_{\rm ph}$ for mixed H/He accretion models at the four ignition depths: $y_b = 3 \times 10^8$, $5 \times 10^8$, $1 \times 10^9$, and $5 \times 10^9 \textrm{ g cm}^{-2}$. From top to bottom, the panels correspond to the four different treatments of convection during the hydrostatic burst rise: the Ledoux criterion (our baseline treatment), the Schwarzschild criterion, predictive mixing, and CPM (all with $\alpha_{\rm sc}=0$). 

At the shallowest ignition depth (y3e8), all convection schemes yield winds composed almost entirely of the accreted material. The only slight exception is the Schwarzschild criterion, which has some mixing between the burning layers and the shallower depths, resulting in the emergence of small amounts of intermediate-mass nuclei (e.g., \isotope[20]{Ne} and \isotope[24]{Mg}).  

For the y5e8 models, the differences between the convection schemes become more pronounced. The Ledoux criterion again yields winds composed mostly of the accreted material.  By contrast, the three other schemes show significant enrichment of intermediate-mass $\alpha$-capture nuclei extending from the Si--Ca region. The Schwarzschild criterion and predictive mixing show the most while CPM shows slightly less. Nuclei such as ${}^{28}\mathrm{Si}$, ${}^{32}\mathrm{S}$, ${}^{36}\mathrm{Ar}$, and ${}^{40}\mathrm{Ca}$ appear prominently, and trace amounts of heavier elements such as ${}^{44}\mathrm{Ti}$ and ${}^{48}\mathrm{Cr}$ are also present.

For the y1e9 models, the Ledoux criterion continues to show limited enrichment, with the wind still dominated by the accretion composition. In the three other convection schemes, however, heavier elements such \isotope[40]{Ca}, \isotope[44]{Ti}, and \isotope[48]{Cr} dominate the composition. Iron-peak elements such as \isotope[52]{Fe} and \isotope[56]{Fe} are also present, demonstrating that the burning has progressed to the final pathways of the \texttt{approx21.net} network.  

At the deepest ignition depth (y5e9), the Ledoux criterion finally shows enrichment of intermediate-mass $\alpha$-capture nuclei. The remaining three schemes continue the trend toward heavier compositions with increasing $y_b$, and are dominated by \isotope[56]{Ni}.

\begin{figure}[t!]
    \centering
    \includegraphics[scale=1]{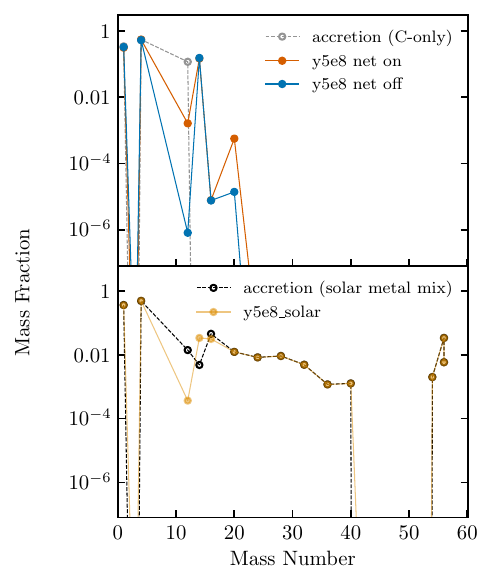}
    \caption{Composition of the wind at the photosphere ($r=r_{\rm ph}$) at the time of maximum $r_{\rm ph}$ for  mixed H/He accretion with $y_b=5\times 10^8 \textrm{ g cm}^{-2}$.  The top panel shows results with nuclear burning turned on and turned off (net on and net off, respectively) during the wind phase. The bottom panel shows the accretion composition and the wind composition assuming a realistic solar abundance accretion rather than the \isotope[12]{C}-only metal accretion of our baseline models. }
    \label{fig:mass_frac_vs_mass_num_solar}
\end{figure}

Overall we see that the choice of convection scheme has a significant impact on the wind composition. The Ledoux criterion consistently yields winds dominated by the accretion composition, while the Schwarzschild criterion, predictive mixing, and CPM all yield substantially ash-enriched winds (though not for the y3e8 models). These results, as well as the $\alpha_{\rm sc}$ results, underscore the critical role of convective mixing physics in determining the wind composition and thus the potential spectroscopic signatures of PRE bursts.

\subsection{Impact of Ongoing Burning and  Metallicity}\label{subsec:impact_burning_and_solar}

Our baseline calculations allow for ongoing nuclear burning during the wind phase and assume \isotope[12]{C} is the only accreted metal in the mixed H/He accretion models.  As explained in Section~\ref{subsec:ignition depths}, the \isotope[12]{C}-only model should be sufficient because the CNO abundances adjust to the equilibrium ratio of $^{14}$O and $^{15}$O in the hot CNO cycle prior to ignition, as noted by \citetalias{Guichandut:2023}.

In Figure \ref{fig:mass_frac_vs_mass_num_solar} we explore how changing each of these approaches impacts the wind composition (for $\alpha_{\rm sc}=0$). The top panel compares the y5e8 model for mixed H/He accretion with nuclear reactions turned on to one with them turned off during the wind phase. Although the abundance profiles are identical at wind onset, they differ shortly thereafter, with the net-on case yielding enhanced abundances of \isotope[12]{C} and \isotope[20]{Ne}. This difference is  likely due to slight differences in the wind-base depth $y_{\rm wb}(t)$, arising from variations in the luminosity evolution between the net-on and net-off cases. It is not due to heavier elements synthesized during the wind phase, as such products are formed too deep to be ejected by the wind.

The bottom panel shows results assuming an accretion composition consisting of a realistic solar abundance pattern based on the results of  \cite{2009M&PSA..72.5154L}. We find that the composition of the wind closely matches the accretion composition just like in the \isotope[12]{C}-only y5e8 model (see Figure~\ref{fig:mass_frac_vs_mass_num_asc}).  This suggests that the \isotope[12]{C}-only model is sufficient for accurately capturing the burst rise and the subsequent wind phase.

\section{Summary and Conclusions}
\label{sec:summary and conclusions}

We carried out hydrodynamic \texttt{MESA} simulations of PRE bursts that explore how ignition depth $y_{\rm b}$, accretion composition, and the treatment of convection influence the structure, evolution, and composition of a burst's radiation-driven winds. Our study extends previous work by modeling a range of ignition depths for both pure He and mixed H/He accretion, while also comparing several prescriptions for convection that modify the extent of compositional mixing during the hydrostatic burst rise.

We find that deeper ignitions produce more powerful, longer lasting winds with somewhat larger maximum photospheric radii $r_{\rm ph}$ and peak mass-loss rates $\dot{M}_{\rm w}$ (Figure~\ref{fig:wind_properties_at_photosphere}).  For a given $y_{\rm b}$, the pure He and mixed H/He accretion models show relatively similar wind structure and evolution, although not composition. This is because in both cases the ignition layer is composed of nearly pure He (in all our models the H depletion depth $y_{\rm d} < y_{\rm b}$) and therefore their burst energetics and radiative-luminosity evolution $L(t)$ are similar. One notable difference is that for a given $y_{\rm b}$, the winds of mixed H/He accretion bursts last somewhat longer, especially for $y_{\rm b} \gtrsim 10^9 \textrm{ g cm}^{-2}$. This is because the H in the outer layers of the atmosphere increases the opacity which lowers the Eddington luminosity $L_{\rm Edd}$ and prolongs the super-Eddington phase $L(t) > L_{\rm Edd}$.

The wind composition is highly dependent on $y_{\rm b}$ and the treatement of convection (Figures~\ref{fig:mass_frac_vs_mass_num_asc} and \ref{fig:mass_frac_vs_mass_num_all}). The larger $y_{\rm b}$ is, the more enriched the wind is with freshly synthesized ashes of nuclear burning.  The reason is two fold.  First, bursts with larger $y_{\rm b}$ have more extended convection zones (smaller minimum $y_{\rm c}$; Figure~\ref{fig:nw_fig6}) and thus more effectively  transport ashes into the shallower layers ejected by the wind.  Second, the larger $y_{\rm b}$ is the further up the $\alpha$-capture chain the burning reaches during the $\approx 1 \textrm{ s}$ long vigorous convective phase that precedes the wind; bursts with $y_{\rm b} \lesssim 5\times10^8\textrm{ g cm}^{-2}$ have only synthesized up to intermediate mass elements by this time, while bursts with $y_{\rm b} \gtrsim 10^9\textrm{ g cm}^{-2}$ have already synthesized Fe-peak elements. 

Even though the wind is launched after the convective phase has effectively ended, the composition of the ash-stratified layers left behind by convection and ejected by the wind depends on the treatment of convective mixing and boundaries during the burst rise.  In order to explore this dependence, we considered three different values of the semiconvection efficiency parameter ($\alpha_{\rm sc}=0,\,0.1,\,0.5$) and four different prescriptions for evaluating the location of convective boundaries available in \texttt{MESA}: the Ledoux criterion (our baseline presription), the Schwarzschild criterion, predictive mixing, and convective premixing (CPM). For the shallowest ignition depth we considered ($y_{\rm b}=3\times10^8\textrm{ g cm}^{-2}$), nearly all treatments of convection result in winds composed primarily of the accreted material, with only trace amounts of freshly-synthesized ashes (the exception is the Schwarzschild criterion, which includes substantial amounts of \isotope[20]{Ne} and some \isotope[24]{Mg}).  However, for $y_{\rm b}\ge 5 \times10^8\textrm{ g cm}^{-2}$, we find that the winds are highly enriched in ashes of burning for the pure He models and for the mixed H/He models at larger values of $\alpha_{\rm sc}$ and in three out of the four convective boundary prescriptions.  The Ledoux criterion is the exception and requires either $\alpha_{\rm sc}=0.5$ or very deep igntion ($y_{\rm b}\gtrsim 10^9\textrm{ g cm}^{-2}$). In general, we find that shallower bursts can eject substantial amounts of intermediate mass $\alpha$-capture elements such as \isotope[20]{Ne}, \isotope[24]{Mg}, \isotope[28]{Si}, and \isotope[32]{S}, while deeper bursts can eject substantial amounts of \isotope[40]{Ca}, \isotope[44]{Ti}, \isotope[48]{Cr}, and \isotope[56]{Ni}.  While this overall trend applies to both pure He and mixed H/He accretion models (Figure~\ref{fig:mass_frac_vs_mass_num_asc}), the former eject ashes even for the Ledoux criterion with $\alpha_{\rm sc}=0$.

Overall, our results show that the wind composition of PRE bursts is not simply a function of ignition depth and accretion composition but also depends critically on the efficiency of convective mixing and the location of convective boundaries. The abundance of heavy nuclei in the wind photosphere is thus a sensitive probe of not just the ignition conditions but also the convective evolution during the burst rise. 

Continued advances in modeling nuclear burning, multidimensional convection, radiation-driven outflows, and radiation transport will further improve the predictive power of PRE burst models.  Particularly useful would be models that carry out time-dependent burst and wind calculations in full general relativity, relax the assumption of local thermodynamic equilibrium and the diffusion approximation in the outer parts of the wind, and use composition-dependent opacities that account for bound-free and bound-bound transitions and Compton scattering. Such improvements would allow for a more complete understanding of PRE bursts, help explain the spectral features observed by {\it NICER}, and might help inform PRE-based measurements of NS radii.

\par\vspace{10pt}
This work was supported by NSF award No. 2107218. We thank Ian J. M. Crossfield for granting us access to the KU ExoLab research group's high-performance server, which enabled us to run a large portion of our \texttt{MESA} simulations.

\vspace{5mm}

\software{This work made use of the Python libraries 
$astropy$ \citep{2022ApJ.astropy},
$NumPy$ \citep{Harris_2020},
$SciPy$ \citep{Virtanen_2020_SciPy},
and $Matplotlib$ \citep{hunter_2007_matplotlib}.
The $py\_mesa\_reader$ package \citep{josiah_schwab_2024_13697200} was used to process {\tt MESA} output files}; \texttt{MESA} (version 23.05.1; Paxton et al. \citeyear{Paxton:11},
\citeyear{Paxton:13},
\citeyear{Paxton:15},
\citeyear{Paxton:18},
\citeyear{Paxton:19},  \citealt {2023ApJS..265...15J}). This work also made use of analysis scripts adapted from \citetalias{Guichandut:2023}.

\bibliography{resources}{}
\bibliographystyle{aasjournal}

\end{document}